# Brain temperature: what it means and what it can do for (cognitive) neuroscientists

**David Papo**[1]*

[1]*Center for Biomedical Technology, Universidad Politécnica de Madrid, Madrid, Spain*
*Correspondence: papodav@gmail.com

The effects of temperature on various aspects of neural activity from single cell to neural circuit level have long been known. However, how temperature affects the system-level of activity typical of experiments using non-invasive imaging techniques, such as magnetic brain imaging of electroencephalography, where neither its direct measurement nor its manipulation are possible, is essentially unknown. Starting from its basic physical definition, we discuss possible ways in which temperature may be used both as a parameter controlling the evolution of other variables through which brain activity is observed, and as a collective variable describing brain activity. On the one hand, temperature represents a key control parameter of brain phase space navigation. On the other hand, temperature is a quantitative measure of the relationship between spontaneous and evoked brain activity, which can be used to describe how brain activity deviates from thermodynamic equilibrium. These two aspects are further illustrated in the case of learning-related brain activity, which is shown to be reducible to a purely thermally guided phenomenon. The phenomenological similarity between brain activity and amorphous materials suggests a characterization of plasticity of the former in terms of the well-studied temperature and thermal history dependence of the latter, and of individual differences in learning capabilities as material-specific properties. Finally, methods to extract a temperature from experimental data are reviewed, from which the whole brain's thermodynamics can then be reconstructed.

**Keywords: temperature; specific heat; free energy; fluctuation-dissipation theorem; symmetry; disorder; fragility; scaling; multifractals; equilibrium; cognitive neuroscience; learning; brain activity; resting state**

## 1. INTRODUCTION

The brain is a dissipative out-of-equilibrium biophysical system, subject to energy, entropy, and information flows across its boundaries. Thus, in principle, it should be possible to describe its activity in terms of thermodynamic variables, e.g. (internal or free) energy, pressure, or temperature [1,2]. Furthermore, inasmuch as cognitive function can be thought of as an emergent property of spontaneous brain activity, it should be possible to characterize cognitive function in terms of these same variables [3].

The brain's energy consumption, ability to do work and the physical conditions under which this can be performed efficiently have been studied extensively [4]. At experimentally relevant time scales, the brain functions at approximately constant pressure. However, other thermodynamic variables such as free energy and temperature undergo important fluctuations. While the significance of free energy to functional brain activity has been explored at length [5], that of temperature remains to be understood.

Many important physical properties, viz. electrical resistivity, viscosity, and chemical reactions rates, generally show marked temperature dependence. It is therefore not surprising that temperature should modulate brain functioning.

Here, we adopt the viewpoint of a cognitive neuroscientist recording brain activity with an electroencephalogram or with functional magnetic resonance, wherein the level of neural activity (noninvasively) observed is essentially macroscopic, and temperature can neither be measured nor manipulated in a direct way.

We illustrate various physical meanings of temperature, and review ways to extract it from data obtained through system-level brain recordings, and to use it to characterize the generic properties of spontaneous brain activity as well as the neural activity associated with cognitive function.

## 2. EFFECTS OF TEMPERATURE ON MICRO AND MESOSCOPIC LEVELS OF NEURAL ACTIVITY

That temperature has profound effects on a wide range of parameters of neural activity at various scales has long been known [6].

At the cell level, ionic currents, membrane potential, input resistance, action potential amplitude, duration and propagation, and synaptic transmission have all been shown to be affected by temperature variations [6-10]. For instance, the ratio between potassium and sodium conductances, altering basic membrane properties, shows temperature dependence [8]. Cooling is associated with neuron depolarization and increases in input resistance [9,11], so that less current is needed to alter the potential, bringing the cells closer to the spiking threshold. Cooling also decreases the conduction of action potentials along axons [12], increases neuronal excitability [8,9] and the latency of excitatory postsynaptic potentials, prolonging their time-course [8]. Relatively minor changes in brain temperature such as those typically associated with physical exercise or fever can modify the amplitude of synaptic potentials in the hippocampus [13,14].

At mesoscopic scales of neural activity, temperature turns out to be a control parameter steering network activity toward different functional regimes. Slow and fast rhythms observed *in vivo* [15,16] and *in vitro* [17] are both sensitive to temperature. Temperature affects the duration, frequency and firing rate of activated states during slow frequency oscillations. Furthermore, cooling affects the ability to end activated states, possibly as a result of a decreased ability to recruit activity-dependent potassium channels thought to contribute to ending up states [18]. Furthermore, while increasing temperature results in enhanced high frequency synchronization in the hippocampus [19], cooling decreases it [20,21], and can therefore be used to reduce epileptic activity [22,23]. Finally, cooling also reduces metabolic processes [24], and has been used as a way to silence cortical areas to study their function [25]. Temperature also has a substantial effect on chemical reaction rates [26], and affects the blood oxygen saturation level by changing haemoglobin affinity for oxygen [27].

At all levels, from single cell to mesoscopic, temperature can *directly* be observed and manipulated. On the other hand, at the system-level of typical non-invasive studies of brain activity, temperature can only *indirectly* be estimated. Its definition and role become less intuitive, and its understanding cannot dispense with a theoretical foundation.

## 3. FROM PHYSICS TO THE BRAIN

Temperature is a physical quantity that measures the mean kinetic energy of the vibrational motions of matter's particles. Its role is to control the transfer of energy between the system and other ones to which it is thermally coupled. Temperature is an intensive property, i.e. it is shared by all the constituents of a system, and independent of system size. Together with potential energies of particles, and other types of particle energy in equilibrium with these, it contributes to the total internal energy within a substance.

*Temperature is defined as the inverse of the entropy variation $\Delta S$ with respect to a variation of the energy $\Delta E$, at fixed volume*

$$-\frac{1}{T} = \frac{\partial S}{\partial E}|_{V,N} \qquad (1)$$

The inverse temperature $\beta = 1/T$ can be seen as a Lagrangian multiplier reflecting the maximization of entropy with respect to internal energy in an isolated system. *Inverse temperature is*, in essence, *the cost, in entropy, of buying energy from the rest of the world* [28]. At low temperatures, a given variation of energy $\Delta E$ results in a large entropy variation $\Delta S$. The system has few excited states and is relatively ordered; a change of energy leads to the activation of many states and thus to a large change in the number of excited states, quantified by entropy. High temperature corresponds to low sensitivity of entropy to variations in energy: the system is excited and disordered.

*Temperature quantifies the amplitude of the fluctuations of a system's physical variables around their expected values* and is a measure of the relative probability that the system possesses a given energy [29]. In a closed system at equilibrium, the probability of a microscopic state $s$ having energy $E(s)$ is given by the Boltzmann distribution $P(s) \propto e^{-\beta E(s)}$, where $\beta$ is a Lagrangian multiplier fixing the value of the energy $E$. This distribution has maximum entropy for a given average energy $\langle E \rangle$.

### 3.1. CONTROLLING BRAIN (THERMO)DYNAMICS

A natural way to understand the role of temperature from a system-level perspective and using non-invasive neuroimaging techniques is to model the brain as a thermodynamical system.

If one assumes the brain to be a closed system exchanging energy with its environment at constant temperature $T$, then activity would evolve towards a state minimizing the free energy $F \equiv E - TS$ where $E$ is the internal energy, and $S$ is proportional to the number of configurations of the system for a given energy $E$.

*The temperature $T$ can be seen as a parameter controlling the balance between order and disorder.* At high temperatures, the entropic term $-TS$ prevails, and the system evolves towards a more disordered state. However, at low temperatures, the energy $E$ is the leading term and the system tends to order so as to minimize its internal energy state.

As the energetic and entropic terms become equal, the system approaches a phase transition. If, as a critical value of the temperature $T_C$ is approached, where both phases have the same free energy, quantities such as spatial or temporal correlations, the heat capacity $C_V = \partial \langle E \rangle / \partial T$ ( i.e. the ratio of the amount of average heat energy transferred to an object to the resulting increase in temperature of the object) or the susceptibility $\chi$, describing the response to an applied field, diverge, and scale as $(T_C - T)^{-\alpha}$, the system is said to undergo a second order phase transition, between phases with different symmetries.

Because resting brain fluctuations generically exhibit scaling properties, it has been suggested that the brain is a system living near the critical point of a second order phase transition [30-32]. In this canonical context, *temperature functions as a control parameter* determining qualitatively different regimes of response ranges. In particular, at criticality, neural networks maximize their dynamic range, i.e. the range of stimuli to which they can respond [33-34].

Similarly, brain activity descriptions in terms of connectivity networks between different regions can be mapped onto a standard thermodynamic system [35] and endowed with an order parameter and a potential. Many topological networks properties can be understood in terms of the low-temperature behaviour of the system [36], and topological phase transitions can be found as temperature is varied [37].

*Temperature can also be understood as controlling the relative importance of cost and available resources* [36]. In fact, in the infinite-temperature extreme, all configurations are equally probable, so that the system does not distinguish between cheap and expensive states. In the zero-temperature regime, the system is forced to severe optimization: only the least costly configuration can be formed and the system occupies the states with lowest energy.

#### 3.1.1. Shaping the phase space and visiting it at temperature speed

Moving from statistics to dynamics, we can represent brain activity as a search for minima on the surface of the entropy production rate $(dS_i/dt)$ in the configuration space of the kinetic variables [1]. The kinetic free energy balance takes the form

$$\frac{dF}{dt} = -\int_V div\ j_F\ dV\ -T\frac{d_iS}{dt} \quad (2)$$

where is $dF/dt$ is the free energy production rate, the second term represents the free energy flow, and $-T(d_iS/dt)$ the rate of free energy dissipation within the system. The system evolves through a series of regressions to temporary entropy production minima alternating with fluctuations which introduce new internal constraints and open new channels for regression. Learning and reasoning can qualitatively be thought of respectively as the temporary storage of free energy within the system and as the nucleation by fluctuation of metastable internal states and the associated increases in entropy production [1]. In this process, *temperature controls the rate of dissipation within the regression-nucleation path*. When the dissipation, or the total increase in entropy, is large, time asymmetry is self-evident. Thus, *temperature controls the price, in terms of entropy lost to dissipation, paid for the break-down of time-translational invariance*.

In general, one can think of brain activity as the motion on a high-dimensional landscape where each point represents a possible microstate of the system, with valleys corresponding to separated regions of flow and their associated attractors, and barriers between them to hills and saddles [38].

Whereas at high temperatures the system explores the whole landscape, at low enough temperatures the dynamics boils down to two processes: a fast relaxation toward local minima via a diffusion process, and a slow activated process in which the system overcomes barriers toward other minima, slowing the evolution of the system [39].

A convenient way of quantitatively characterizing brain motion within the landscape is to describe it as the dynamics of a macroscopic particle subject to a viscous friction, changing with a time scale $\tau_m$, and to an additive random force $\eta(t)$, with time scale $\tau_\eta$ [40,41].

This dynamic representation makes explicit that *temperature controls not only transport processes, such as diffusivity, but also the characteristic times and velocities of the system*. For instance, in the simple case of a Brownian particle, where the random force has fast-vanishing δ-correlated Gaussian fluctuations (i.e. $\tau_\eta \ll \tau_m$), $\tau_m$ scales with inverse temperature as

$$\beta \propto \frac{\tau_m}{D_m} \quad (3)$$

*Temperature also controls the characteristic times of phase space navigation*. For instance, a Brownian particle trapped into a potential well of depth $h \gg T$, and rarely jumping out into one of the neighbouring wells, escapes with a rate $k$ following the *Arrhenius law*

$$k \propto exp\left(-\frac{\Delta E}{T}\right) \quad (4)$$

where $\Delta E$ denotes the threshold energy for activation, so that, taking $k \equiv 1/\tau_{Esc}$, the dwelling time scales with inverse temperature as

$$\beta \propto \frac{1}{\Delta E} log(\tau_{Esc}) \quad (5)$$

### 3.2. TEMPERATURE AS A BRIDGE FROM RESTING TO TASK-RELATED BRAIN ACTIVITY

A *bona fide* temperature ought to reflect heat flows and thermalization, i.e. how fluctuations relax to states in which the values of macroscopic quantities are stationary, universal with respect to differing initial conditions, and predictable [42].

The notion of temperature is intimately related to that of *equilibrium*. Operationally, equilibrium is defined by the *zeroth law* of thermodynamics, which states that if two systems are in thermal equilibrium with a third one, they must be in thermal equilibrium with each other. Thermometers can then be used to establish whether two systems will remain in thermal equilibrium when brought in contact.

The *zeroth law* allows using thermal equilibrium as an equivalence relationship on the set of thermally equilibrated systems, inducing a partition into subsets in mutual equilibrium. Temperature maps these subsets onto real numbers, with ordering and continuity properties. Thus, provided an appropriate thermometer can be devised, *temperature can be used as a macroscopic collective variable describing the system, through which value its different subparts can be sorted*.

#### 3.2.1. Thermometers and the fluctuation–dissipation theorem

A thermometer is a device, e.g. an oscillator, which when coupled to a given observable $X$, feels on the one hand its fluctuations, measured by the two-time autocorrelation function $C_X(t,t') = \langle X(t)X(t')\rangle$, in the absence of perturbations, and on the other hand, the result of its own action on the system, proportional to the response function $R_X(t-t')$, i.e. how $X$ responds at time $t$ to a small perturbation at time $t'$ [43]. In the Fourier domain, these two latter quantities are respectively replaced by the power spectrum $G(\omega)$ and by the response function $R_X(\omega)$.

For a system at equilibrium, these two opposing effects give the correct energy, i.e. the one predicted by *equipartition theorem*, for every thermometer and observable, only if correlations and responses associated with any observable are proportional

$$T = \frac{\partial C_X(t,t')/\partial t}{R_X(t-t')} = \chi(t,t')/C_X(t,t') \quad (6)$$

where $\chi(t,t') = \int_{t'}^{t} R_X(t,\tau)\ d\tau$ is the integrated response, or equivalently, in the Fourier domain

$$T = \frac{\omega G_X(\omega)}{2\ Im\ R_X(\omega)} \quad (7)$$

The *fluctuation–dissipation theorem* (FDT) ensures precisely that, for a system at equilibrium, *the temperature T of the bath with which the system is in equilibrium represents the ratio between the response to an external field conjugate to some observable and the corresponding autocorrelation function in the unperturbed system* [44].

Applied to brain activity, the FDT establishes a *substantial relation between spontaneous and stimulus-related brain activity*. Thus, at least in principle, *brain responses evoked by a stimulus can be understood through a suitable observation of the correlation properties of brain fluctuations at rest without applying the stimulus* [3]. *The temperature quantifies exactly this relation.*

Interestingly, if the observable $X$ is the local energy of a signal, the FDT ensures that $T$ quantifies the relation between energy fluctuations and the heat capacity $C_V$:

$$T^2 \propto \frac{\langle E - \langle E \rangle \rangle^2}{C_V} \quad (8)$$

Insofar as $C_V$ measures the number of states accessible per temperature unit [45], *temperature regulates the rate at which the system makes microstates available as a function of fluctuations in its energy levels*, consistent with the basic definition of temperature of (1).

#### 3.2.1.1. From Brownian particle to the real brain

The equilibrium formulation of the FDT would imply an additive random force $\eta(t)$ having fast-vanishing $\delta$-correlated Gaussian fluctuations [44]. Activity would hop without memory from a given configuration to some other, and in the long time limit, the temporal autocorrelation of macroscopic velocity fluctuations would exponentially decay with time.

However, brain activity cannot be considered at equilibrium or even close to it, and brain fluctuations are neither Gaussian nor $\delta$-correlated [46-48].

In non-equilibrium systems, the FDT is not expected to hold. The equilibrium temperature $T$ no longer completely characterizes probability distributions for the system's degrees of freedom, so that for instance, the velocity and position distributions of particles are no longer specified. Only fast fluctuations thermalize to the bath temperature $T$. Slow modes, on the contrary, do not, and the direction of heat flows is characterized by an *effective temperature* $T_{eff} > T$ [49]. $T_{eff}$ is, in essence, what a thermometer responding on the time scale on which the system slowly reverts to equilibrium would measure [42]. For such systems, a generalized FDT can be written as

$$\frac{T}{X(t,t')} = \frac{\partial C(t,t')/\partial t}{R(t,t')} \quad (9)$$

where $X(t,t')$ is the fluctuation-dissipation ratio (FDR), and the ordinary FDT is recovered for $X = 1$ [50]. The time-dependent *effective temperature* $T_{eff}(t,t') \propto T/X(t,t')$ allows quantifying the distance to equilibrium, and the extent to which the FDT is violated, at a given scale of activity.

Interestingly, both $T$ and $T_{eff}$ retain *dynamical information about the system*. At thermal equilibrium, the temperature $T$ is brought about by work exchanged through thermal fluctuations and viscous dissipation, two dynamical properties. In non-equilibrium stationary states, $T_{eff} \propto W/(d_iS/dt)$, where $W$ is the work per unit time done on the system by an external force, while the entropy production rate $d_iS/dt$ reflects the average state space contraction rate [51].

At the time scales typical of cognitive processes such as learning or thinking, brain fluctuations are generically characterized by relaxation times considerably slower than exponential, reflecting the divergence of the microscopic time scale [47,52-54], weak ergodicity breaking [45,55], a regime where all possible states are still accessible, but some require exceedingly long times to visit [56], and *aging*, i.e. time-dependent correlations [45].

For systems presenting these properties, generalized forms of the FDT are likely to apply, with $T_{eff}$ in (9), and the corresponding phase space navigation times and velocities, respectively in (4) and (5), taking a specific, typically non-exponential, functional form in accordance [42,57,58].

Contrary to equilibrium fluctuations, which are time homogeneous and for which both the correlation $C$ and the response function $R$ depend on $\tau = t - t_w$ elapsed from the instant $t_w$ at which a field is applied, for aging systems these quantities separately depend on both $t_w$ and $t$, so that the *age* $t_w$ becomes a relevant time scale. The corresponding effective temperature is given by $T_{eff}(t,t_w) \propto T/X(t,t_w)$.

#### 3.2.1.2. Multithermalization, heterogeneity and sorting by temperature

The brain is a driven nonequilibrium system which generically responds to changing external fields with a series of avalanches spanning a broad range of scales [59], corresponding to different thermalization rates.

In an equilibrium system, any thermometer coupled to a part of the system reads the same temperature [43]. In out-of-equilibrium systems such as the brain this exchange happens at widely different timescales simultaneously, reflecting the inherent multiscale character of brain activity. A system can be at equilibrium on one scale and out of equilibrium on another, or may even be in equilibrium but show scale-dependent properties [50,60]. Each timescale may be associated with its own FDR, containing information about the relaxation of the process and $T_{eff}$ [61]. This allows *understanding the relationship between spontaneous and stimulus-induced brain activity at each scale, and the extent to which each scale of brain activity deviates from equilibrium conditions, produces entropy etc.*

When taking into account its spatial extension, the brain can be considered as a heterogeneous system, as at any given time, different regions in the brain relax at different rates. $T_{eff}$ can be used to estimate the degree of dynamical heterogeneity, i.e. of spatiotemporal fluctuations in the local dynamical behaviour, of the whole system. This can be done by calculating the *dynamic susceptibility* $\chi_T(t) = \partial \langle C(t) \rangle / \partial T$, with temperature as the perturbing field [62].

## 4. TEMPERATURE AND COGNITION

Once brain activity is endowed with a thermodynamic characterization and its dynamics with a landscape representation, it is intuitive to see how temperature may act as a parameter controlling processes such as learning, thinking or reasoning, which can all naturally be modelled as search processes within a complex landscape.

We briefly illustrate for the case of learning ways in which temperature can be used to study brain activity associated with cognitive function.

### 4.1. LEARNING AS A THERMALLY-GUIDED PROCESS

In statistical mechanical [63] and machine learning [64] models, learning is a stochastic dynamics through which the system seeks to reach a given location within the landscape. Much as in the brain, the dynamics essentially boils down to adjusting the neural connectivity pattern until a target pattern is found [65]. Given the role of temperature both in shaping the landscape and in determining transport properties and moments of the dynamics within it, learning is naturally characterized as a thermally-guided process.

The temperature $T = 1/\beta$, which in essence gives the variance of the noise in the learning process, facilitates thermally activated jumps of free energy barriers, and can be used to account for costs and benefits of network rewiring, as quantified by an energy function $E_t$ of the error of a given training set [63].

Temperature can be used as an optimization tool, as for instance in *simulated annealing* [66], an algorithm in which thermally activated jumps of free energy barriers are induced by introducing stochastic imprecision which is and then gradually reduced to locally fine-tune computation [39].

#### 4.1.1. Annealing, brain plasticity and learning

Viewing learning-related brain activity as a thermal process akin to a heat treatment, and particularly as an annealing process, has some interesting implications. In annealing, a metal or a glass is heated until it reaches a temperature at which it is too hard to deform but soft enough for stresses to relax, and finally slowly cooled at a rate proportional to the heat capacity.

Plasticity and learning-related brain processes can then be portrayed as resulting from the interaction between external fields and temperature conditions, forcing the system into a condition in which it can modify its structure.

This framework allows using known results from the physics of materials to investigate the temperature dependence of a brain system successfully undergoing learning, the thermal conditions learning may be facilitated or hampered, or to predict whether some system has an ability to learn. For example, the *time-temperature superposition principle* [67] indicates that, *ceteris paribus*, stress accumulated by a material relaxes faster at higher temperatures.

Implicit within this perspective is the representation of the brain as a material of some kind. This has at least two important aspects. On the one hand, fundamental properties of materials, such as elasticity and plasticity, viscosity or fragility, are temperature-dependent. On the other hand, the temperature dependence of properties such as plasticity is material-specific, and this may allow understanding inter-individual differences both in spontaneous and task-related brain activity.

##### 4.1.1.1. Brain matter and amorphous materials

Some of its generic properties, viz. slow non-exponential relaxation times and aging, make spontaneous brain activity in many ways comparable to amorphous materials such as glasses, under certain thermal conditions regimes [68]. These materials are characterized by a of lack long-range order typical of liquids, but their inability to flow makes them dynamically similar to solids.

Amorphous materials are naturally endowed with a temperature-based operational definition. A glass can be obtained from most materials in their liquid phase by cooling them fast enough to avoid crystallization, so that it remains in a disordered metastable liquid state, until it effectively becomes solid, below the so-called glass transition temperature $T_g$ [69].

The aging process in glasses is also explicitly connected to the notion of temperature and thermal treatment. In amorphous materials, aging designates the ultraslow relaxation to equilibrium and the gradual increase in viscosity undergone as the temperature is kept fixed after cooling. The system however does not reach the equilibrium value at the new fixed temperature within experimental times [70]. In fact, the older the system the slower it relaxes.

Interestingly, this phenomenology is consistent with a recently proposed model of synaptic dynamics, where each synapse has a cascade of states with different plasticity levels, connected by metaplastic transitions, which quantify the ability of the system to undergo further deformation, and memory traces present power-law decay [71].

As the system *ages*, the number of available configurational states diminishes [72] and the corresponding effective temperature $T_{eff}$ higher than the equilibrium temperature $T$. External fields, on the contrary, force the system out of equilibrium, *rejuvenating* the system [73] in a manner similar to the way external stimuli affect scaling exponents in brain activity [74]. $T_{eff}$ and available configurations are related by the expression

$$\frac{1}{T_{eff}(t,t_w)} = \frac{\partial \Sigma(f,T)}{\partial f} \qquad (10)$$

where $\Sigma(f,T)$ is the configuration entropy, which quantifies the number of states with a given free energy at a given temperature, so that $T_{eff}$ *counts the number of metastable states of the system in the same way as $T$ reflects the number of microstates in a system at equilibrium* [75].

The structure and physical properties of a glass depend not only on the temperature, but also on the time $t_w$ after preparation, on the thermal processes it underwent during its formation and the order in which they were applied [39]. For instance, the slower the cooling, the longer the time available for configurational sampling at each temperature, hence the colder the system can become before falling out of equilibrium. Thus, the temperature at which the system effectively becomes solid increases with cooling rate [76]. Transferring this to brain characterization indicates that the plasticity range may depend on the thermal schedule followed during learning, and not just on the intrinsic properties of the brain system undergoing it. The role of thermal history in amorphous materials can in particular be used to interpret the critical importance of learning schedules, e.g. of distributing practice over time, to the strength of learning, observed by cognitive scientists [77].

### 4.1.1.1.1. Fragility in the brain

The rate at which the structural relaxation time $\tau_\alpha$ or the shear viscosity $\eta$ increases for $T \to T_g$ helps classifying glass-forming materials. Materials are said to be *strong* if $\tau_\alpha$ and $\eta$ show the Arrhenius behaviour defined by (4), and *fragile* if they increase much faster with decreasing temperature [69,78].

In the typical viscosity range of materials used for blowing, a fragile liquid cools faster and in a smaller temperature range than hard glasses, leaving much less leeway for deformation [79].

One stimulating advance would then be to define a brain fragility index based on some function of brain activity, for instance the degree to which the system deviates from equilibrium and ergodicity [80]. Such an index could help characterizing not only a neural system's ability to learn but also the time-windows within which learning can occur, as well as the schedule under which it may be optimized. It would also be interesting to identify the conditions for the transition from strong to fragile behaviour [78,81].

## 5. TEMPERATURE IN REAL DATA

### 5.1. ESTIMATING TEMPERATURE

Functionally induced brain temperature changes and the associated spatio-temporal scales can be estimated using the model of brain temperature proposed in [82,83].

Heat in the brain is produced by oxygen consumption and removed by blood flow. The amount of locally *generated heat* $Q_r^+$ [$\mu mol \cdot g^{-1} \cdot min^{-1}$] is proportional to the regional oxygen metabolic rate $rCMRO_2$ weighted by the difference between the enthalpy generated by the reaction between oxygen and glucose $\Delta H^0$ and the energy used to release oxygen from haemoglobin $\Delta H_b$:

$$Q_r^+ = (\Delta H^0 - \Delta H_b) \cdot rCMRO_2 \qquad (11)$$

The rate of *heat removal* from brain tissue $Q_r^-$ can be estimated as the product of regional cerebral blood flow $rCBF$ [$ml \cdot g^{-1} \cdot min^{-1}$] and the difference between tissue and arterial temperatures, weighted by blood heat density $\rho_B$ and heat capacity $C_B$

$$Q_r^- = rCBF \cdot \rho_B \cdot C_B \cdot (T - T_{arterial}) \qquad (12)$$

For brain activity at rest, the local *steady state* temperature $T_0$ can be estimated by

$$T_0 = T_{arterial} + \frac{(\Delta H^0 - \Delta H_b)}{\rho_B \cdot C_B} \cdot \frac{rCMRO_2}{rCBF} \qquad (13)$$

where $T_{arterial}$ is arterial inflow temperature [82].

The model estimates in the order of a few millimeters the characteristic length $\Delta$ of regions where temperature changes can be observed, with differences between superficial and deep regions [83].

Functional activity changes the oxygen extraction fraction $OEF = rCMRO_2/rCBF$. Since typically $rCBF > rCMRO_2$, the model predicts that local changes in temperature and in $rCBF$ always have opposite sign [83].

Changes in global $CBF$ induce a temperature dynamics with a relaxation time $t_T = C_{tissue}/(rCBF \cdot \rho_B \cdot C_B)$. Estimates of $t_T \sim 40$–$60$ s [83] indicate that for $t < t_T$, below the vascular response scale, measurements are out of equilibrium, $T$ is not well defined, and $T_{eff}$ should be estimated.

The model [82,83] allows inferring from functional magnetic resonance data that functional stimulation can induce local brain temperature fluctuations of up to ±1°C with respect to resting temperature, by locally changing the balance between metabolic heat production and heat removal by blood flow. These values are consistent with *indirect* temperature estimations using the temperature dependence of the magnetic resonance signal's frequency [84-86].

The potential impact of temperature modulations on functional brain activity is significant. Given a temperature effect on blood oxygen saturation levels of several percent/1°C [26], and an estimated average brain van't Hoff temperature coefficient $Q_{10}$ (the factor by which a reaction rate increases for 10°C increases) of 2,3 [27], the observed temperature fluctuations may lead to sizeable changes in blood oxygen saturation levels and to >2% variations in chemical reaction rates.

Importantly, the model provides quantitative indications on *steady state* temperature modulations, and the precision with which these can be evaluated, but say little on the fluctuations that these may undergo.

### 5.2. DIRECT MEASUREMENT

Experimental neuroscientists typically ought to extract information from time series of some aspect of brain activated gather with some (often non-invasive) recording device, e.g. electro or magnetoencephalogram, or functional magnetic resonance.

Temperature could simply be measured by applying (1) to e.g. the local amplitude of the neural signal recorded by a standard non-invasive technique or, more accurately and whenever possible, by confronting evoked and spontaneous brain activity. This can be done by plotting $\chi(t,t')$ against $C_X(t,t')$ [87,88]. For equilibrium systems, this yields a straight line with slope $-1/T$. Out-of-equilibrium systems can have a more complex $\chi - C_X$ relationship, depending on the particular properties of the system. For instance, multiscaleness and aging lead to a nonlinear $\chi - C_X$ plot [60], and a corresponding spectrum of slopes.

The experimental analysis of the dependence of $T_{eff}(t, t_w)$ on $t$ and $t_w$ helps distinguishing between different models of aging as the FDT violations are model dependent [89]. The aging properties can be studied by monitoring the time evolution of correlation and response functions. In the former case, one compares the configuration of spontaneous activity at $t_w$ and $\tau + t_w$. In the latter, one follows the evolution of the linear response to a perturbation applied at $t_w$. Deviations from the FDT can be estimated by plotting the integrated linear response $\chi(t, t_w)$ against the correlation for fixed $t_w$, varying $t$ between $t_w$ and infinity. As $t_w$ diverges, $\chi$ converges to an integral function of $T_{eff}$ [90]. For weak ergodicity breaking, asymptotically the FDR should depend on time only through the correlation function $X(t, t') = X(C(t, t'))$, so that the integrated response $\chi(t, t') = \chi(C(t, t'))$ completely describes the ageing dynamics [91].

The estimated $T_{eff}$ can then be used to quantify the degree of dynamical heterogeneity of the whole system by calculating the *dynamic susceptibility* $\chi_T(t)$ with an appropriate ansatz [62].

### 5.3. GENERALIZED TEMPERATURES

Comparing spontaneous and stimulus-induced activity may not always be possible experimentally. Nonetheless, quantities that can be interpreted as temperatures by analogy to thermodynamics can still be derived from experimental data.

So far, we have proposed a thermodynamic characterization of the brain as a macroscopic system in terms of an intensive variable, the temperature, the quantification of which allows reconstructing the thermodynamic functions of the system, e.g. the heat capacity $C_V$. In a formally equivalent way, one may try to derive a temperature from the statistical properties of the fluctuations of some aspect of brain activity recorded during the course of an experiment.

#### 5.3.1. Probability distributions

Brain activity is naturally described as a superposition of dynamics at different time scales, each corresponding to a different thermalization process. A global temperature can be defined by considering the probability distribution functions of the moments of some aspect of brain activity at various scales.

In analogy with fully developed turbulent flows, brain activity could be considered as a structure where kinetic energy is injected at large scales generates large vortices. As scales decrease, smaller vortices are generated at increasing speed, until energy is ultimately dissipated through viscous interactions.

In [92] it was proposed that within the so called inertial range, between the scales at which the energy is respectively injected and dissipated, $P(\delta v_r)$, the probability distribution of longitudinal velocity differences over a distance $r$ is characterized by large intermittent fluctuations leading to a non-Gaussian distribution, with Gaussian and heavy-tailed non Gaussian fluctuations respectively at large and small scales. To account for this transition, it was proposed that the overall distribution of $P(\delta v_r)$ results from the superposition of different Gaussian distributed velocity scales conditioned to a given energy dissipation rate $\varepsilon_r$, weighted by the probability of each dissipation rate

$$P(\delta v_r) = \int_0^\infty P(\varepsilon_r) P(\delta v_r | \varepsilon_r) \, dv \varepsilon_r \quad (14)$$

$P(\varepsilon_r)$ is in essence the probability of observing a characteristic velocity $\sigma_1$ at scale $r_1$, given a value $\sigma_2$ at scale $r_2$. *The variation of the logarithm of this quantity with the logarithm of the scale ratio $(\ln \sigma_1/\sigma_2)$ at equilibrium can be interpreted as an inverse temperature*. The former term plays the role of internal energy and the latter of entropy of equation (1). This temperature-like quantity is conserved across scales when the whole cascade is at equilibrium. Furthermore, it depends on the Reynolds number and therefore, as expected of a temperature, it reflects the relative strength of inertial to viscous forces in the flow [93].

#### 5.3.2. Superstatistics and time-varying temperatures

A temperature can be extracted from experimental data even when considering the non-equilibrium character of brain activity. One way to do so it to think of the brain as a system in a non-equilibrium steady state, where its many subsystems are temporarily in local equilibrium [94].

Each subsystem can then be associated with an approximately constant intensive variable $\beta$, taken from a probability distribution $f(\beta)$. Because the whole system is not in equilibrium, each subsystem may in fact have a different effective temperature. Altogether, $\beta$ is a slowly varying stochastic process, and the whole system can be regarded as a field with a spatial correlation length of the order of the subsystem size $L$, and a temporal correlation length $T$ much larger than the local relaxation time to equilibrium [94]. For $t \gg T$, the system is associated with an effective Boltzmann factor, where $e^{-\beta E}$ is weighted by the probability distribution $f(\beta)$ which can be regarded as the *superstatistics* i.e. the *statistics of the statistics $e^{-\beta E}$* associated with the subsystems of the system.

Interestingly, if brain activity is understood as a Brownian particle [40,41], $\beta$ is exactly the inverse temperature of (3).

The local relaxation times, superstatistical time scales $T$, and corresponding temperature values together with the size of the subsystems $L$ can all be estimated from experimental time series [95-97]. Relaxation times can be determined by studying the respective autocorrelation function, the time $T$ by looking at the time scale at which observed fluctuations have Gaussian distribution, as measured by the kurtosis. Finally, the temperature $\beta(t)$ can be determined by the variance of the local Gaussians, and the empirical distribution $f(\beta)$ is the histogram of $\beta(t)$ for all the integrated time interval [95].

#### 5.3.3. Local scaling properties

A temperature can also be derived from microscopic information about the *local scaling properties* of the system. This can be done by considering observed time series as dynamical trajectories visiting the phase space of the underlying dynamical system which geometrical and dynamical properties they allow reconstructing, under rather general conditions [98,99], and by identifying some of its properties to a temperature.

While experimental evidence does not seem to militate in favour of a turbulence-like scenario à la Castaing for brain activity [100], it points nonetheless to a complex attractor with a very irregular multifractal form [53,54,100], characterized by a hierarchy of dimensions exhibiting self-similar scaling properties and a corresponding spectrum of scaling exponents [101,102].

To characterize the geometrical properties of the attractor from the data, the experimental measure is covered with boxes $i$ of size $\epsilon$ which the trajectory visits with probability $p_i(\epsilon)$, so that the time series is transformed into the sequence of cells visited by the system. In the long time limit, the natural measure $\mu(.)$ quantifying, the fraction of time an orbit spends in any given region of the state space (and therefore the attractor's inhomogeneity) is finally reconstructed.

The *thermodynamic formalism* [103] allows expressing the properties of a dynamical system in terms of thermodynamic-like functions [104]. This involves a structural analogue of the equilibrium partition function $Z(\beta) = \sum_i exp(-\beta E_i)$ of statistical mechanics. Rather than microscopic states and configurations, the *structural partition function* $Z(\epsilon, q) \equiv \sum_i p_i^q$, where $p_i$ is the probability of visiting a part of the phase space of size $\epsilon$, weights the relative density of different parts of the attractor [103].

Temperature is identified with the order $q$ of the *generalized dimensions* [105]

$$D_q = \lim_{\epsilon \to 0} \frac{1}{q-1} \frac{ln[Z(\epsilon, q)]}{ln(\epsilon)} \quad (15)$$

which correspond to the scaling exponents of the $qth$ moments of the natural measure $\mu(.)$, and can directly be measured from experimental data [102,106]. The equivalence between inverse temperature $\beta$ and $q$ is made explicit by the relations

$$Z(\beta) \equiv \sum_i exp(-\beta E_i) = \sum_i exp[q\, ln(p_i)] \quad (16)$$

With similar analogies, it is possible to derive other thermodynamical functions. For example, the free energy is identified with the scaling exponent of the dynamical partition function $Z(\epsilon, q) \sim \epsilon^{-\tau(q)}$, and the non-analyticities of moments of $\tau(q)$ for certain values of $q$, e.g. of the heat capacity $C_V = \partial^2 \tau(q)/\partial q^2$ [106], can be thought of as equivalents to thermodynamical phase transitions.

Finally, the temperature-free energy relationship for a given region-of-interest can be obtained by averaging the values of $\tau(q)$ obtained from the recordings across that region and repeating the regression for a range of values of $q$.

A more authentically thermodynamic characterization has recently been proposed for fractal time series [107]. A Hamiltonian is written for the observed dynamics and the whole system's thermodynamic functions reconstructed in such a way that the time series statistics is completely governed by an effective temperature quantifying the scattering of the fractal dimension of the time series.

## 6. CONCLUSIONS

We discussed various definitions of temperature, together with the roles it can play and the ways in which it can be quantified at the macroscopic level of brain activity of standard system-level non-invasive neuroimaging recordings, at which it cannot directly be measured in a trivial way.

Once derived experimentally, effective temperatures can identify the FDT governing brain activity and therefore the non-equilibrium regime at which it is working. This information could be used to describe brain activity not through its level of activation but through a measure of disequilibrium and its heterogeneity at various spatial and temporal scales.

We showed various ways in which temperature can be treated not only as a *control* parameter, steering brain activity to various regimes, but also as an *order* parameter i.e. as a collective variable directly describing it. One could in principle observe how temperature varies during a given experimental condition, e.g. the execution of a cognitive task, and then how phase transitions may occur, using temperature as a control parameter and some other property of neural activity as the order parameter.

More generally, the assessment of temperature and thermal history enables both a dynamical description of brain activity, and a complete characterization of its thermodynamics, affording neuroscientists a description of the object of their investigations with a sound physical basis.